\documentclass[final,5p,times,sort&compress]{elsarticle}
\usepackage{fixltx2e}
\usepackage{epstopdf}
\usepackage{amsmath,amsfonts,amssymb}
\usepackage[colorlinks=true,linkcolor=magenta]{hyperref}
\usepackage{multirow}
\usepackage{xspace}
\makeatletter
\DeclareRobustCommand\onedot{\futurelet\@let@token\@onedot}
\newcommand{\@onedot}{\ifx\@let@token.\else.\null\fi\xspace}
\newcommand{\eg}{{e.g}\onedot\xspace} 

\newcommand{\ie}{{i.e}\onedot\xspace}

\newcommand{\etc}{{etc}\onedot\xspace}

\newcommand{\etal}{\textit{et al}\onedot\xspace}

\makeatother
\makeatletter
\def\blfootnote{\xdef\@thefnmark{}\@footnotetext}
\makeatother
\journal{Journal of Alloys and Compounds}

\begin{document}
\begin{frontmatter}
\title{Theoretical Study of Defects  in Cu$_3$SbSe$_4$: Search for Optimum Dopants for Enhancing Thermoelectric Properties}

\author{Dat T. Do\corref{cor1}}
\ead{dodat@msu.edu}
\ead[url]{http://www.msu.edu/~dodat}
\author{S. D. Mahanti\corref{cor2}}
\ead{mahanti@pa.msu.edu}
\ead[url]{http://www.pa.msu.edu/~mahanti}
\address{Department of Physics and Astronomy, Michigan State University, East Lansing, MI 48824, USA}
\cortext[cor1]{Corresponding author}
\cortext[cor2]{Principal corresponding author}
\date{\today}
\hyphenation{tetra-hedral-ly iso-struc-tural pseudo-poten-tail com-pounds}

\begin{abstract}

Cu$_3$SbSe$_4$ is a promising thermoelectric material due to high thermopower ($>400\ \mu$V/K) at 300K and higher. Although it has a simple crystal structure derived from zinc blende structure, previous work has shown that the physics of band gap formation is quite subtle due to the importance of active lone pair (5$s^2$) of Sb and the non-local exchange interaction between these and Se 5$p$ electrons. Since for any application of semiconductors understanding the properties of defects is essential, we discuss the results of a systematic study of several point defects in Cu$_3$SbSe$_4$ including vacancies and substitutions for each of the components. First principles calculations using density functional theory show that among variety of possible dopants, $p$-type doping can be done by substituting Sb with group $IV$ elements including Sn, Ge, Pb and Ti and $n$-type doping can be done by replacing Cu by Mg, Zn. Doping at the Se site appears to be rather difficult. Electronic structure calculations also suggest that the $p$-type behavior seen in nominally pure Cu$_3$SbSe$_4$ is most likely due to Cu vacancy rather than Se vacancy.
\end{abstract}
\end{frontmatter}
\section{Introduction}
Thermoelectrics (TE) are materials used in thermal-electrical energy conversion. An important physical quantity that controls the performance of a TE is the dimensionless figure of merit $ZT=\sigma S^2 T/\kappa$, where $\sigma$ is the electrical conductivity, $S$ is the Seebeck coefficient (thermopower) and $\kappa=\kappa_{el}+\kappa_{latt}$ is the thermal conductivity which is the sum of thermal conductivity of electrons ($\kappa_{el}$) and phonons ($\kappa_{latt}$). The efficiency of a TE system increases nonlinearly with ZT and approaches the Carnot efficiency as $ZT\rightarrow\infty$. An important goal of thermoelectric study is to achieve as large value of $ZT$ as possible. This can be achieved by increasing  $\sigma S^2$ (called the power factor) through the engineering of electronic structure\cite{pei_band_engin.2012} and/or manipulating the energy dependence of electron scattering rate, and arresting phonon transport (reducing $\kappa_{latt}$ that usually dominates $\kappa_{el}$ in high performance TEs which are doped semiconductors) by introducing defects of different length scales (point defects, nanoparticles, grain boundaries \etc) and taking advantage of intrinsic lattice anharmonicity\cite{Snyder2008.complex.thermo}.

Since good thermoelectrics are narrow band gap semi\-con\-ductors\cite{sofo_mahan94} their electronic transport properties have to be optimized by carefully controlling the types of defects and their concentrations. This optimization is necessary because the two electronic transport quantities controlling the power factor are contra-indicatory \ie increasing $\sigma$ reduces $S$ or vice versa, the well-known Pizarenko relation\cite{ioffe57}. Thus it is important to investigate what types of defects can be inserted into a host semiconductor, what are their formation energies and how they affect the electronic structure of the host and ultimately charge and energy transport. The ideal case is when the defect acts as a simple dopant without altering the band structure of the host, the so called rigid band approximation. In many cases this approximation fails and one has to be careful in analyzing experimental results.\cite{malsoon12}

In this paper we address the question of defects in a promising thermoelectric compound, Cu$_3$SbSe$_4$ (Se4). Cu$_3$SbSe$_4$ has the Famatinite\cite{datahandbook} crystal structure (sometimes referred as 3-1-4 materials) which belongs to the tetrahedrally coordinated (diamond-like) class of compounds whose common characteristics are the average of four valence electrons per atom and tetrahedral coordination of all the atoms. This class of materials covers a wide range of semiconductors including the well-known group IV elements, III-V (such as GaAs and AlAs), II-VI (\eg ZnS, ZnSe), 1-1-2 (such as CuInS$_2$, CuInSe$_2$, AgGaSe$_2$) and 2-1-3 (such as Cu$_2$Ge$S_3$, Cu$_2$SnSe$_3$, Ag$_2$GeSe$_3$) compounds \cite{datahandbook}.
However, only a limited number of these compounds have attracted attention for thermoelectric applications; Cu$_3$SbSe$_4$ and its isostructural analogs are among them. One of the rationale to investigate this and similar compounds is that they do not contain Pb (poisonous) and Te (volatile and expensive) as in well-known thermoelectrics such as PbTe, (GeTe)
$_{0.85}$(AgSbTe$_2$)$_{0.15}$ (commonly known as TAGS) with large $ZT$ values\cite{Snyder2008.complex.thermo}.

In a recent study Skoug \etal\cite{Skoug.doping.2011} have shown that Se4 as synthesized is a $p$-type TE with rather large $S$ values ($>400\ \mu$V/K at T~300K). In a later work\cite{skoug11.solidsolution} they also showed that by optimizing carrier concentration through partial substitution of Sb by Ge and making solid solution of Se4 and its isostructural analog Cu$_3$SbS$_4$ to reduce the lattice thermal conductivity, they could improve the $ZT$ values. They found that in Cu$_3$Sb$_{0.97}$Ge$_{0.03}$Se$_{2.8}$S$_{1.2}$, $ZT\sim1$ could be achieved at $T\sim700K$, comparable to the well-known optimized PbTe\cite{Snyder2008.complex.thermo}. If one can reduce the thermal conductivity of Se4 further through nanoparticle insertion (as is done in PbTe-SrTe\cite{biswas_high-performance_2012}) one can get larger $ZT$ values. As regards $n$-type Se4, it turns out to be extremely difficult to dope Se4 $n$-type.\cite{skoug_thesis_2011} 

Our previous work on Se4\cite{do_cusbse.2012} has shown that, in spite of the simple diamond-like crystal structure, the electronic structure of Cu$_3$SbSe$_4$ near the band gap region is rather subtle. From a simple valence count (Cu$^{1+}$, Se$^{2-}$) one expects Sb to be 5+ in this compound. In contrast, we found that Sb is in nearly 3+ configuration with its 5$s$ states (lone pair) fully occupied. However the 5$s^2$ lone-pair electrons of Sb strongly hybridize with the $p$ states of Se and split one band out of the twelve Se $p$ bands (per formula unit). This split-off hybridized band becomes the lowest conduction band. Our calculations predict that Se4 when doped can be a good $p$-type TE with large $S$, in agreement with experiment.\cite{do_cusbse.2012} We found that $n$-doped Se4 has values of $S$ in acceptable range for TE application provided $n$-type compounds can be synthesized. We also showed that the band gap of Se4 can be tuned by using complete isovalent substitutions\cite{do_bond_band2014}.

 Despite the fact that Se4 has the potential to be a promising TE there are very few experimental studies on these compounds \cite{Skoug.doping.2011,skoug11.solidsolution,li_Bi_Sb.2013,wei_Cu_deficiency.2014,wei_Sn_Sb.2014} and, to the best of our knowledge, no theoretical work on defects in this system is available in the literature. In the present work we address this deficiency by calculating the formation energies of a large number of neutral dopants (isovalent and non-isovalent) using first principles electronic structure calculations. In addition we  investigate the effect of defects on the electronic structure and potential thermoelectric properties.

The arrangement of the paper is as follows: in Sec.~\ref{comput}, we describe the method we have used to calculate the formation energies of defects and then discuss briefly the computational methods used. We present our results and discussion in Sec.~\ref{result}. The summary and the highlights of our work are given in Sec.~\ref{sec.sum}.
\section{\label{comput}Method to calculate formation energy and computational details}
The defect system is modeled by creating a defect in a periodic supercell. Most of the calculations have been done starting from a 64 atom (24 Cu, 12 Sb and 32 Se) supercell of Se4. In some cases, a 128 atom supercell was used to see what happens to the physical properties for lower concentrations of defects. To study the energetics of defects, we follow the work of Zhang and Northrup\cite{zhang_northrup.91} where the formation energy $\Delta{}E_f(X)$ of a defect $X$ of charge $q$ is given by:
\begin{align}
	\Delta E_f (X)&=E_{tot} (X)-E_{tot} (0)+\sum_An_A E_A+n_X E_X+q\epsilon_{VBM}\notag\\
	&+\sum_An_A\mu_A+n\mu_X+q\epsilon_F,\label{eqn.eform}
\end{align} 
where $E_{tot} (X)$ and $E_{tot} (0)$ are the total energies of the system with and without the defect $X$ respectively, $E_A$ and $E_X$ are the ground state energies (per atom) of the solid state of the host atom $A$ ($A$= Cu, Sb, or Se) and the impurity $X$ accordingly. $n$ is the number of atoms transferred from the reservoir (pure solid) into the host system; $n>0$ if an atom is transferred from the host material to the reservoir and $n<0$ for the reverse situation. $\epsilon_{VBM}$ is the valence band maximum energy of the host and $\epsilon_F$ is the Fermi energy  referenced to $\epsilon_{VBM}$. $\mu_A$ and $\mu_X$ are the chemical potentials of $A$ and $X$ respectively (as measured from the atomic chemical potential of the elemental solid at $T=0K$) . One can see that the defect formation energy depends not only on the atomic energies (in the solid state) but also on the atomic chemical potentials which are determined by the experimental conditions. For example one can control the chemical potential by growing samples under Cu rich or poor conditions.  Under equilibrium growth conditions, the range of chemical potentials is however bounded by thermodynamic conditions which require\cite{zhang_northrup.91}
\begin{align}
	\mu_A&\le0,\mu_X\le0\\
	3\mu_{Cu}+\mu_{Sb}+4\mu_{Se}&=\Delta{}E_f (Cu_3 SbSe_4)\\
	m\mu_A+l\mu_X&<\Delta{}E_f (A_mX_l )
\end{align}
In this study, only neutral defects ($q$=0) are considered and the atomic chemical potentials are assumed to be 0. Thus, Eq.~\ref{eqn.eform} reduces to:
\begin{equation}
	\Delta{}E_f (X)=E_{tot} (X)-E_{tot} (0)+\sum_An_A E_A+n_X E_X
\end{equation}
For example, the formation energy of a Cu vacancy is given by
\begin{equation}
\Delta{}E_f (vac_{Cu} )=E_{tot} (vac_{Cu} )-E_{tot} (0)+E_{Cu},
\end{equation}
and the formation energy of a Ni substituting a Cu is given by
\begin{equation}
\Delta{}E_f (Ni_{Cu} )=E_{tot} (Ni_{Cu} )-E_{tot} (0)+E_{Cu}-E_{Ni}
\end{equation}
Value of $\Delta{}E_f$ determines how likely it is to create a defect in the host material. Positive formation energy means it costs energy while negative means creating defects releases energy.
\begin{figure*}
\centering
\includegraphics[width=.8\textwidth]{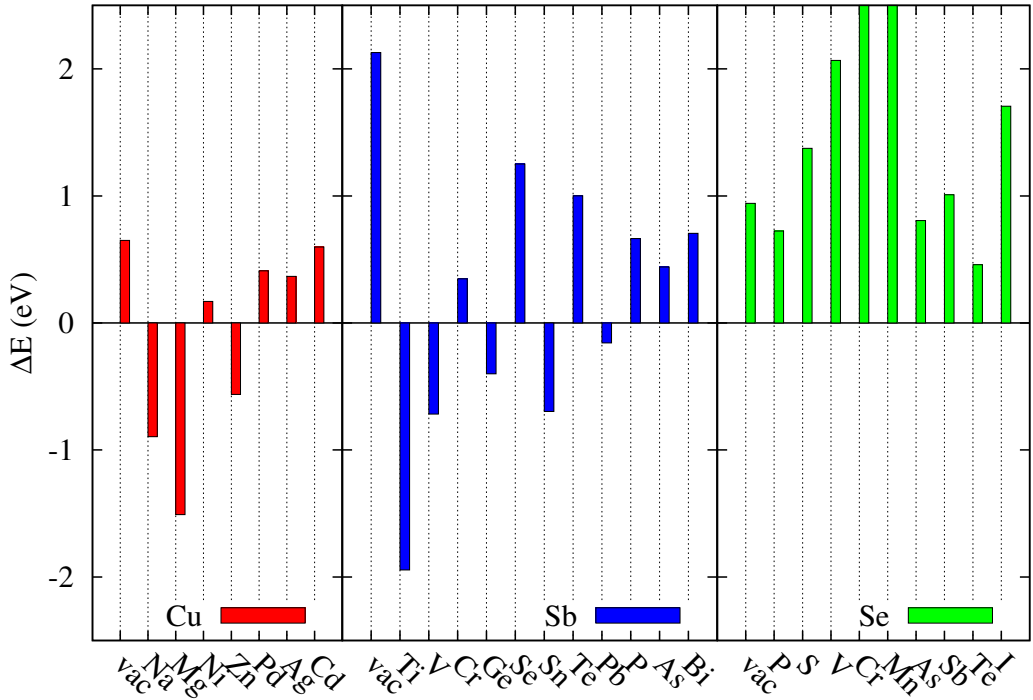}
\caption{\label{fig.formen}Formation energy of substitution defects for each component in Cu$_3$SbSe$_4$}
\end{figure*}

Total energy calculations are performed using pseudo-poten\-tial, projector augmented wave (PAW) method\cite{bloch94,kresse99} within the generalized gradient approximation (GGA) with Perdew-Burke-Ernzerhoff (PBE) exchange-correlation potential\cite{pbe} as implemented in Vienna Atomic Simulation Package (VASP) \cite{vasp1,vasp2,vasp3}. Energetics within GGA are known to be reliable, however, it usually underestimates the band gap of semiconductors and insulators (some cases even predicted to be metal in GGA)\cite{martin_book}. In fact, in earlier studies\cite{do_cusbse.2012,do_bond_band2014} we found that GGA gives a negative band gap in Se4 and the observed semiconducting gap  could be obtained either after incorporating non-local exchange partially (through HSE06 model\cite{hse06:1,hse06:2,hse06:3} or by introducing a large effective value of the Coulomb repulsion ($U$) among the Cu $d$ electrons within the GGA+$U$ model\cite{dudarev98}. If one is interested in knowing the formation energies of charged defects ($q\neq0$ in Eq.~\ref{eqn.eform}), then a correct value of the band gap (hence the Fermi energy $\epsilon_F$) is essential. For neutral defects, however, GGA is reasonably good and we have used this approximation to calculate the formation energies. In the total energy calculations, the energy cutoff and convergence criteria are set to 400 eV and 1E-4 eV respectively and the Brillouin zone are sampled using Monkhorst-Pack scheme\cite{monkhorst76}. In all the calculations, the atomic structures were allowed to relax.

In addition to the calculations of the formation energies we have also looked at the changes in the host electronic structure brought about by the defects using the supercell model\cite{defects}. In this case however one has to have the correct band gap. Since computing electronic structure for the defective systems (supercells with 64 or 128 atoms) using nonlocal approximations (HSE06) is extremely time-consuming and computationally expensive, we have used the GGA+$U$ model (local) to calculate the band structure and the band gap. Since the $d$ bands  of Cu are full and affect the band structure near the band gap (associated with p states of Se and s states of Sb) indirectly through hybridization one can tune this hybridization by controlling the position of Cu d bands though U.  In our previous study of the pure compound we had to use a rather large effective value of $U$ ($\sim$15 eV instead of the usual 3-4 eV) to get the correct band gap\cite{do_cusbse.2012,do_bond_band2014}. This value of $U$ is certainly unphysical and should be used carefully in defect calculations. This procedure of using a large parametric value of $U$ to get the correct band gap is somewhat equivalent to the phenomenological approaches suggested earlier by Christensen in the context of LMTO calculations \cite{chistensen_correction.1984} and Segev \etal in the context of pseudo potential calculations\cite{segev_self_corrections.2007} to obtain correct band gaps and band structures of binary semiconductors GaAs, GaN, and InN. In these calculations by introducing additional parameters in the potential (pseudo potential) one adjusts the band gap without affecting the structural parameters by more than a few percent. In a similar vain, in our GGA+ effective U calculation which gave the correct band gap, the structural parameters differed by less than 1\% from the GGA values.

\section{\label{result}Results and discussions}
\subsection{\label{result.formationen}Formation energy of defects}
\begin{figure*}
\centering
\includegraphics[width=.7\textwidth]{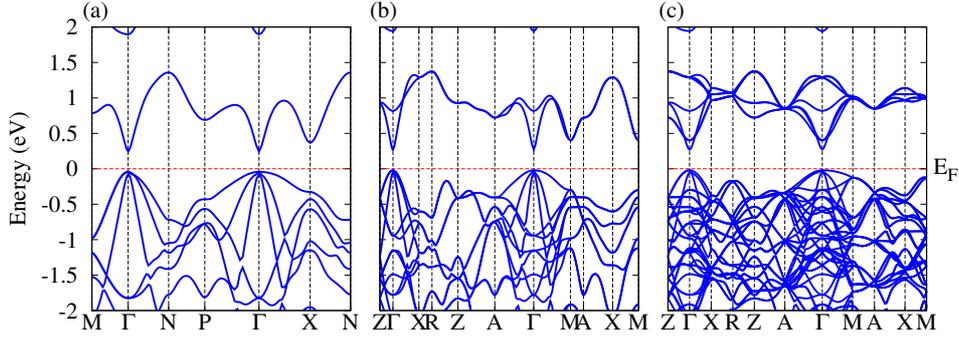}
\caption{\label{fig.scband}Band structures of Cu$_3$SbSe$_4$ in (a) body center tetragonal primitive unit cell (8 atoms/cell), (b) $1\times1\times1$ simple tetragonal cell (16 atoms/cell) and (c) $2\times2\times1$ simple tetragonal cell (64 atoms/cell)}
\end{figure*} 

The results of formation energies are given in Fig.~\ref{fig.formen} for vacancies and different types of substitutional defects. Let us discuss the vacancies first. As we see, the formation energies of vacancies are all positive increasing from Cu (0.65~eV) to Se (0.94~eV) to Sb (2.13 eV); it costs energy to form vacancies at all the atomic sites. Clearly Sb is the most strongly bonded whereas Cu is the least strongly bonded. One can use a simple local (nearest neighbor) bond energy picture to understand this ordering. Since Sb and Cu are bonded to four Se atoms each, using the formation energies of these two atoms we can estimate Sb-Se bond energy as 0.53 eV and Cu-Se bond energy as 0.16 eV. Since Se is bonded to three Cu and one Sb we can estimate its formation energy as $3\times0.16+0.53 = 1.01$ eV. The actual GGA value is 0.94 eV, suggesting that a local bond energy picture caputures the basic physics. Also stronger Sb-Se bond can be understood in terms of additional bonding between Sb 5$s$ (lone pair) and a properly symmetrized linear combination of $p$-orbitals of the four nearest-neighbor Se atoms. In fact, the corresponding anti-bonding combination leads to one out of twelve Se $p$-bands to split off from the rest eleven bands resulting in the opening of a band gap in Se4\cite{do_cusbse.2012}. These results are consistent with experimental observations where Se and Cu vacancies are likely to form in synthesizing process and were attributed to the observed $p$-type behavior of as synthesized Cu$_3$SbSe$_4$ in ref.\cite{Skoug.doping.2011} and ref. \cite{wei_Cu_deficiency.2014} respectively. However, as we will show later when discussing the electronic structure that only Cu vacancy acts as an acceptor whereas Se vacancy does not appear to contribute charge carriers. 

As regards substitutional defects, for each site, we study selective isovalent elements, nominally $n$- and $p$-doping $s$-$p$ elements (to the right and the left of Cu, Sb, Se in the periodic table) and also several transition metal atoms (Ti, V, Cr, Mn \etc). The reason for studying transition metal substitutions is to choose $s^md^n$ instead of $s^mp^n$. By doing that we investigate how different types of atomic orbital affect the materials' electronic structure. As we will show, the changes in the band structure brought about by the defects are fundamentally different between these two types. When we substitute at the Se (anion) site the formation energies are all positive. In contrast, when we substitute at the cation sites we find that formation energies can be both negative (in the range 0,$-2.0$ eV) and positive (in the range 0, 3.5 eV). Substitutions of Na, Mg, and Zn at the Cu site and Ge, Sn, Pb (all $s^2p^2$), Ti ($s^2d^2$), and V ($s^2d^3$) at Sb site give negative formation energies, meaning it is easy to use these atoms as dopants on the respective sites. In experiments, however, only $p$-type doping using group IV substitutions at the Sb site has been reported.\cite{Skoug.doping.2011,wei_Sn_Sb.2014}

\subsection{\label{result.elect_struct}Defect induced electronic structure}

The formation energy analysis discussed above shows how likely a defect can be created in a system. However it cannot confirm whether a defect dopes the system or changes the host band structures. To address this, one has to look at the change in the band structure brought about by the defects. Even though our formation energy calculations shows that there are only a handful number of potential dopants, for completeness, we have considered vacancies and a wide variety of isovalent and non-isovalent defects both at cation (Cu and Sb) and at anion (Se) sites. 

Since all the calculations have been done within a supercell model, we must first understand how the band structure of pure Se4 looks like for the supercell. In Fig.~\ref{fig.scband}, we give the band structures of Se4 obtained in 3 different unit cells, body centered tetragonal (bct with 8 atoms/uc), simple $1\times1\times1$ tetragonal (16 atoms/uc) and $2\times2\times1$ tetragonal (64 atoms/uc) -- the last one is what we have used for defect calculations. The band structures look different due to band folding as the first Brillouin zone becomes smaller in larger unit cells. Also the Brillouin Zone changes, if one focuses on the lowest conduction band (LCB) in the bct unit cell, there is only one band, the split-off Se $p$-Sb $s$ band discussed earlier. When one doubles the unit cell ($1\times1\times1$ tetragonal) the LCB consists of two bands and when we use a $2\times2\times1$ tetragonal cell, LCB contains eight bands. On the other hand, the highest valence band (HVB) remains three-fold degenerate. Band structures of systems with defects should be compared with Fig.~\ref{fig.scband}c. 

Before discussing different defects individually one can make several general observations. (1) Some of the defects preserve the semiconducting gap whereas others either destroy the gap or give rise to localized trap states in the gap; (2) Some of the non-isovalent defects which preserve the band gap and do not modify the band dispersion near the band edges can be treated in a rigid band approximation where as others can not.

\begin{figure}
\centering
\includegraphics[width=.7\columnwidth]{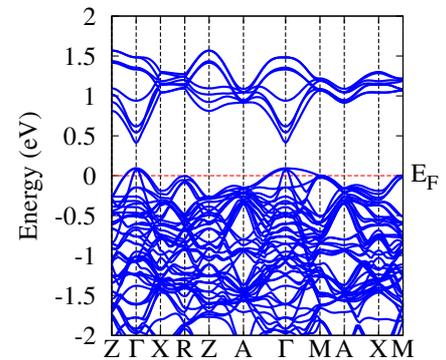}
\caption{\label{fig.cuvac}Band structure of Cu$_{23}$Sb$_8$Se$_{32}$ (one copper vacancy in the $2\times2\times1$ supercell)}
\end{figure} 

\begin{figure*}
\centering
\includegraphics[width=.7\textwidth]{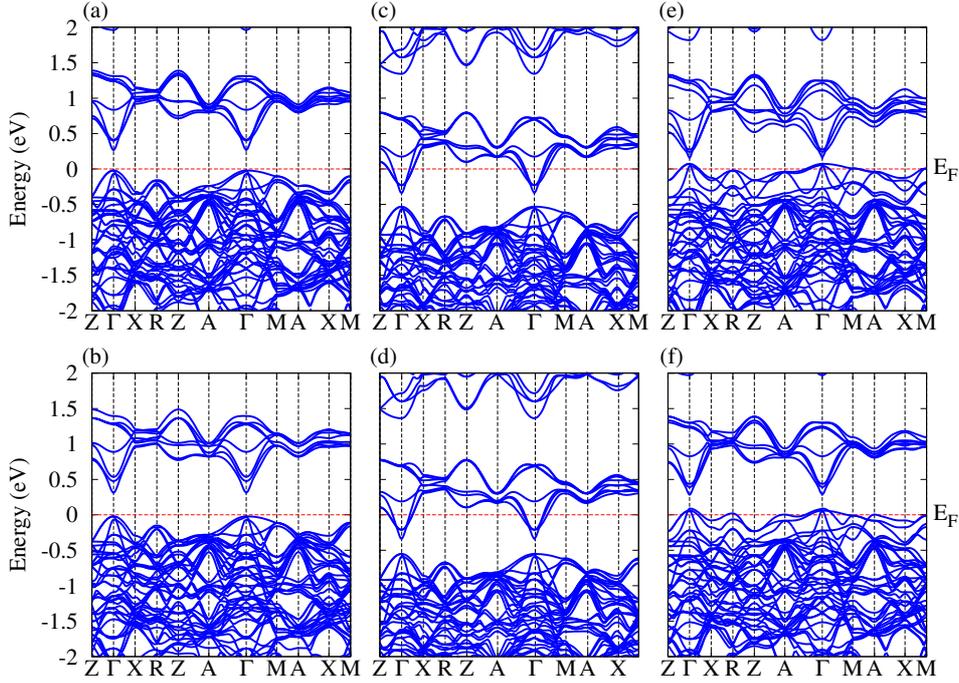}
\caption{\label{fig.cu.band}Band structures of Cu$_3$SbSe$_4$ with Cu substituted with isovalent atoms: (a) Ag$_{Cu}$ and (b) Na$_{Cu}$, divalent atoms: (c) Mg$_{Cu}$ and (d) Zn$_{Cu}$ and nominal $p$-type dopants: (e) Ni$_{Cu}$ and (b) Pd$_{Cu}$; the values of effective $U$ for Ni and Pd defects were chosen to be same as that for Cu.}
\end{figure*}

\subsubsection{\label{result.cu_defect}Defects at the Cu site}

As we have discussed earlier, at the Cu site, substitution of Na, Mg, and Zn have negative formation energies whereas vacancy, substitutions of Ag, Cd, Ni, and Pd have positive formation energies. The band structures for different defects are given in Fig.~\ref{fig.cuvac} (vacancy), and Fig.~\ref{fig.cu.band} (substitution). 

As one sees in Fig.~\ref{fig.cuvac}, one Cu vacancy in the unit cell of volume 1.478E-21 cm$^3$ does not change the band structure much. States near the HVB are hardly affected and there is a small splitting of the LCB. Since there is 1 hole/u.c. the carrier concentration is 6.77E+20 cm$^{-3}$, an order of magnitude larger than seen in experiment (Skoug \etal\cite{Skoug.doping.2011}). For the experimental hole concentration we expect a rigid band picture to be reasonable.  This result suggests that the often observed $p$-type behavior of as prepared Cu$_3$SbSe$_4$ can be ascribed to native Cu vacancies. This has been confirmed in experiment by Wei \etal\cite{wei_Cu_deficiency.2014} in which they control the hole concentration by controlling the Cu deficiency. 

As seen in Fig.~\ref{fig.cu.band}a,b despite belonging to different groups, isovalent impurities (Na, Ag) at the Cu site do not change the band structure or dope the system. This is consistent with our previous work\cite{do_bond_band2014} where we showed that the compound Na$_3$SbSe$_4$ has the same physical origin of the band gap as in Se4. Ag substitution has slightly weaker effect than Na substitution as the latter splits the band degeneracy and ordering of the conduction bands as seen in Fig.~\ref{fig.cu.band}b. This suggest that isovalent substitutions on Cu can be used to reduce thermal conductivity (due to phonon scattering and mass fluctuation) without sacrificing the power factor, which indeed has been done for  the related compound Cu$_3$SbS$_4$\cite{suzumura_s4_2014} where 3\% of Ag$_{Cu}$ was used to further decrease lattice thermal conductivity beside doping on Sb site.

The alkali earth atoms (Mg, Zn, Cd) are divalent impurities and because of the negative formation energies of Mg, Zn these two appear to be excellent $n$-dopants. The band structure corresponding to one extra electron/uc is given in Fig.~\ref{fig.cu.band}c,d. As we can see the HVB and the LCB are hardly affected, just the Fermi energy moves into the conduction band. Again, for these two defects a rigid band picture is quite good. However, as far as we know Mg- and Zn-doped Cu$_3$SbSe$_4$ have not been reported experimentally.

The band structures for Ni and Pd defects are shown in Fig.~\ref{fig.cu.band}e,f. Since we are using the GGA+ effective $U$ model to obtain the band structure we have to choose appropriate effective $U$ values for the Ni and Pd impurities.  Since Ni is in the same row as Cu (3$d$) we can use the same $U$ value (15 eV). For Pd which is in the next row (4$d$) the real $U$ should be smaller by $\sim$2eV compared to Cu. However, in Fig.\ref{fig.cu.band}f, we have used $U$=15 eV, same as in Cu, for Pd as well. From Fig.\ref{fig.cu.band}e,f we see that the states near the valence band top are not affected by the defects, only the band gap changes a little. So we can treat Ni and Pd as good hole-dopants where a rigid band picture is good if one can create a proper synthesizing condition to overcome the positive formation energies.
\subsubsection{\label{result.Sb_defect}Defects at the Sb site}

\begin{figure}
\centering
\includegraphics[width=.7\columnwidth]{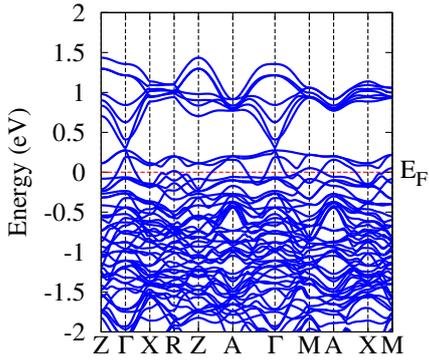}
\caption{\label{fig.sbvac.band}Band structure of Cu$_{24}$Sb$_7$Se$_{32}$ (one Sb vacancy in the $2\times2\times1$ supercell)}
\end{figure} 

When we introduce a vacancy at the Sb site, its effect on the band structure is quite dramatic. The number of conduction bands in the u.c. goes from 8 to 7 (Fig.~\ref{fig.sbvac.band}). This is caused by the absence of Sb 5$s^2$ lone pair at the vacancy site. The $p$-state associated with the Se atoms surrounding the vacancy does not have a Sb lone pair to hybridize with and its energy drops to the valence band region. As a result there is an increase in the number of Se $p$-valence bands by one. Also the loss of one Sb which was contributing three electrons to the valence band results in a $p$-type doping. Unfortunately the perturbation introduced by Sb vacancy is so strong that the band gap vanishes and the resulting system is not expected to be a good thermoelectric. Calculations using a larger supercell ($2\times2\times2$ containing 128 atoms) show that even a smaller Sb vacancy concentration still destroys the band gap. On the other hand, because of the large positive formation energy, it may be difficult to synthesize a system with Sb vacancy. 

\begin{table*}
\caption{\label{tab.sb_nearest}Nearest neighbor distance of the the substitutional defect on Sb}
\centering
\begin{tabular}{lrrrrrrrrr}
\hline\hline
Defect&Pure&P&As&Bi&Te&Sn&Ti&V&Cr\\
\hline
$d$(X-Se) (\AA)&2.56&2.28&2.41&2.68&2.64&2.58&2.42&2.32&2.26\\
\hline\hline
\end{tabular}
\end{table*}

\begin{figure}
\centering
\includegraphics[width=1\columnwidth]{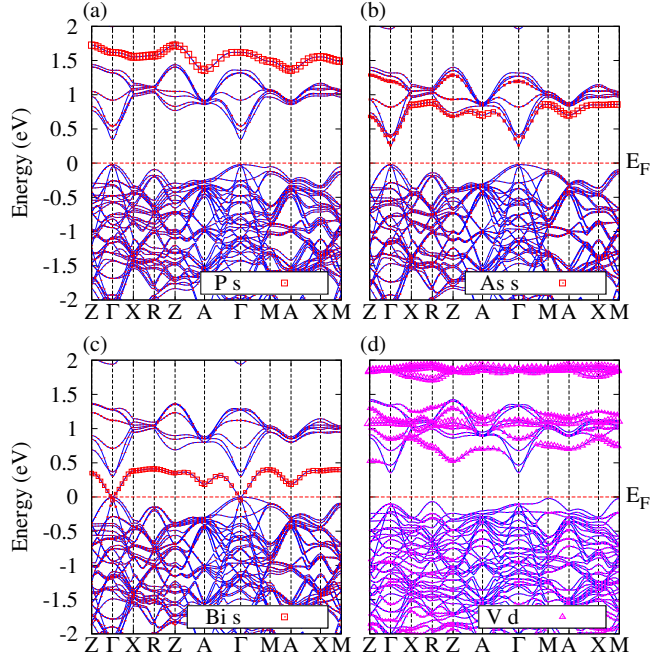}
\caption{\label{fig.sb_iso.band}Band structure of Cu$_3$SbSe$_4$ with isovalent substitutions on Sb site: (a) P$_{Sb}$, (b) As$_{Sb}$, (c) Bi$_{Sb}$, and (d) V$_{Sb}$. The symbols present the orbital characters of the bands.}
\end{figure}

For isovalent impurities (P, As, Bi), the valence bands are not affected much, but the LCB are strongly perturbed. Arsenic changes the conduction band structure the least, removing some degeneracies at the $\Gamma$ point. In contrast, in P one out of the eight lowest conduction bands gets pushed up whereas in Bi one band gets pushed down. In Fig.~\ref{fig.sb_iso.band} we give the band strucutres of iso-valent substitutions on Sb together with the fat-band characters showing that the changes in conduction bands are due to the defect atoms.  This difference can be understood by looking at the energies of the LP states for P, As, Bi ($ns^2$ , $n=3,4,6$ with energies $-19.22$, $-18.92$, $-15.19$ eV) and the differences in the distance between the defect and the neighboring Se atoms for the three defects (Table~\ref{tab.sb_nearest}). The net result is the difference in the strength of the hybridization between the LP and the Se $p$ orbitals. This result is consistent with our earlier calculations in Cu$_3$MSe$_4$, M=P,As,Bi where local (GGA+ effective $U$) and non-local exchange (HSE06) opened up a gap between the topmost valence band and the lowest conduction band in P and As but not in Bi.\cite{do_bond_band2014} The distance between Sb and M and neighboring Se atoms played an essential role in the band structures of Cu$_3$Sb(X)Se$_4$. In the impurity systems local relaxation around the defect controls this distance and also plays an important role in determining the energy spectrum near the Fermi energy. As shown in Table~\ref{tab.sb_nearest}, the distance between defect atoms and the neighboring Se increases when one going from P to Bi, the same trend as shown our previous work\cite{do_bond_band2014}. Our result for Bi is in agreement with experiment on effects of bismuth doping done by Li \etal\cite{li_Bi_Sb.2013} where they found a dramatic decrease of activation gap in resistivity measurement. 

In addition to the iso-valent impurities belonging to the same column as Sb we also investigated the change in band structure when Sb (valence configuration $s^2p^3$) is replaced by V(valence configuration $s^2d^3$). In case of V we expect its $s$-state to be empty (no lone pair at the V site) and $d$-states to contribute new states near the gap. The band structure of V substituting Sb is shown in Fig.~\ref{fig.sb_iso.band}d. V does perturb the top of the valence band much more than P, As, and Bi. It introduces two more potential hole pockets by bringing the maxima at R and M points closer in energy to that at the $\Gamma$ point. The highest valence band shifts from $\Gamma$ to the M point. In addition, one of the eight Se $p$ conduction bands drops down in energy and merges in the valence band and V gives two $d$-bands (e$_g$ symmetry, as shown in Fig.~\ref{fig.sb_iso.band}) to the group of lowest conduction bands (there are nine bands now instead of eight). The five valence electrons of V are just enough to fill all the valence band states, thus preserving the semiconducting behavior. We will discuss later how Cr ($s^2d^4$) instead of V modifies this picture and whether it dopes $n$-type? 
 
\begin{figure}
\centering
\includegraphics[width=\columnwidth]{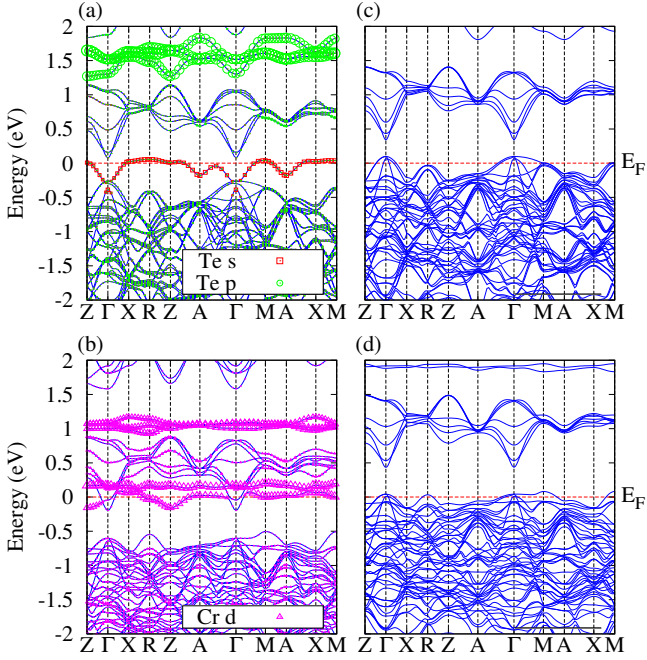}
\caption{\label{fig.sb.band}Band structure of point defects at Sb site with nominal $n$-type dopants: (a) Te$_{Sb}$ and (b) Cr$_{Sb}$ and $p$-type dopants: (c) Sn$_{Sb}$ and (d) Ti$_{Sb}$. The symbols present the orbital characters of the bands.}
\end{figure} 

In order to see whether one can substitute an $n$-type dopant at the Sb site we tried replacing it by Se and Te. We also looked at Cr since it is to the right of V in the periodic table which as we found acted like an isovalent impurity \ie it did not introduce any charge carrier.  The band structures for Te and Cr substituted compounds are shown in Fig.~\ref{fig.sb.band}a,b. Results for Se are similar to Te and are not given. 

We find that instead of acting as an $n$-type, Te acts like a $p$-type dopant. This can be understood by looking at the band structure given in Fig.\ref{fig.sb.band}a where we have also shown the orbital characters of different bands. The three Te $p$-states are shown in green and lie high in energy. However the Se-$p$ states which hybridize with Te LP states ($5s^2$) form the band shown in red and this band splits off from the seven other conduction bands which arise from hybridization of Se-$p$ with the Sb LP states. This can be understood by looking at the difference in the energies of Sb LP ($-16.03$ eV) and Te LP ($-19.12$ eV), the latter hybridizing weakly with Se-$p$ states (energy -10.68 eV). Further more the nearest neighbor (nn) distance between Te and Se is nearly the same as that between Bi and Se in the case of Bi$_{Sb}$ defect (Table~\ref{tab.sb_nearest}), making the physics of the Te substitution similar to that of Bi substitution. The split-off lowest conduction band is indeed half filled but its density of states is such that it acts like a $p$-doped system. This is an example of a defect which is nominally $n$-type but changes the band structure dramatically such that one sees a $p$-type behavior.
The Cr defect behaves as expected. The effects of Cr defect on the band structure are similar to that of V. However, with one extra d electron than V, Cr moves the Fermi level to the bottom of the conduction band making the system $n$-type. With the two flat $e_g$ bands of Cr acting as resonant states, one should expect a large thermopower, similar to thallium impurity in PbTe\cite{Heremans_distorted_dos.2008}.	


Finally we will discuss what happens to the band structure when we replace Sb by a $p$-type dopant ($s^2p^2$ configuration) such as Sn (or Ge, Pb) or by Ti ($s^2d^2$ configuration). The band structures of Sn (Ge and Pb have similar band structures) and Ti defects are shown in Fig.~\ref{fig.sb.band}c,d. We find that Sn is a perfect $p$-type dopant, its LP orbitals ($5s^2$) act just like Sb LP orbitals, the eight conduction band states are slightly perturbed (due to lowering of symmetry) and the top most valence bands are hardly affected. The absence of one valence electron (compared to Sb) leads to a perfect hole-doped system. In fact, Skoug \etal\cite{Skoug.doping.2011} and Wei \etal\cite{wei_Sn_Sb.2014}  obtained their best thermoelectric performance for the Sn doped samples. In contrast to Sn, Ti doping changes both the valence band structure (near the top) and the conduction band structure (seven instead of eight bands due to the absence of LP for Ti). The two $s$ electrons of Ti are used in filling the extra Se $p$ valence band that drops down from the group of eight). Thus the hole count for Ti is same as Sn. It is, however, noteworthy that Ti defect introduces hole pockets at R and M points of the Brilloun Zone in addition to the one at $\Gamma$. Thus one should expect larger thermopower in Ti-doped system compared to the Sn-doped system.
 
\subsubsection{\label{result.Se_defect}Defects at the Se site}
\begin{figure}
\centering
\includegraphics[width=.7\columnwidth]{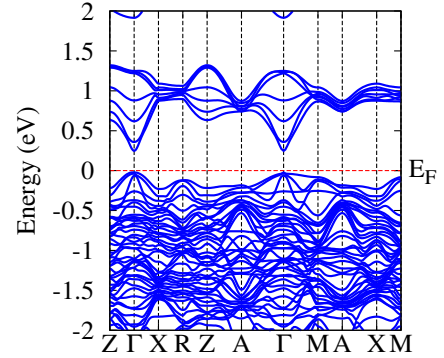}
\caption{\label{fig.sevac.band}Band structure of Cu$_{24}$Sb$_8$Se$_{31}$ (one Se vacancy in the $2\times2\times1$ supercell)}
\end{figure}  

\begin{figure*}
\centering
\includegraphics[width=.7\textwidth]{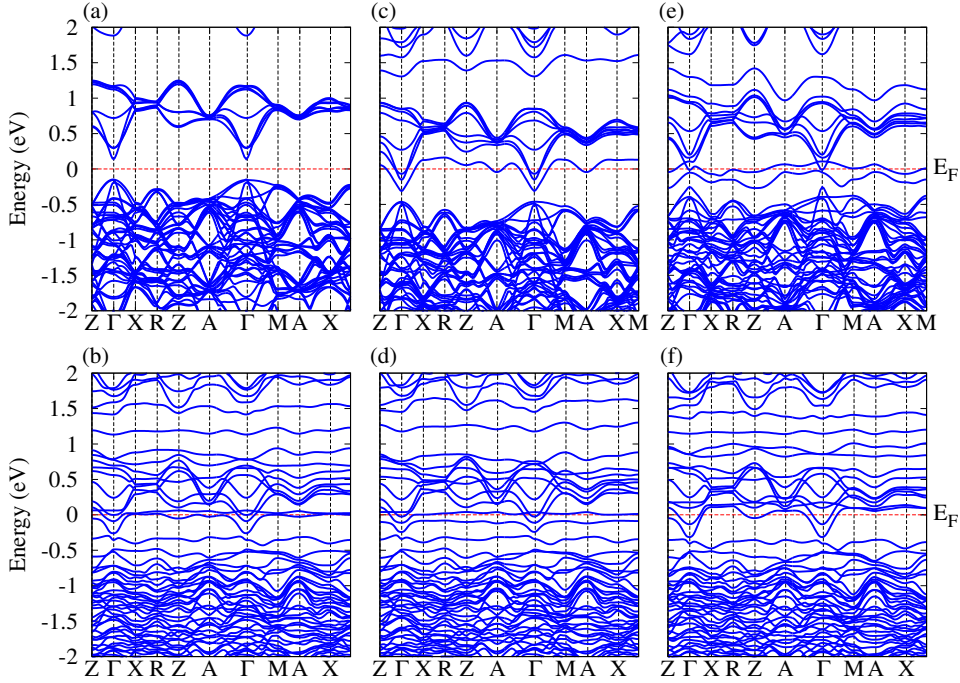}
\caption{\label{fig.se.band}Band structure system with point defects at Se site: (a) S$_{Se}$ and (b) Cr$_{Se}$, (c) I$_{Se}$,(d) Mn$_{Se}$, (e) Sb$_{Se}$ and (f) V$_{Se}$}
\end{figure*} 

The band structure for a Se vacancy (Cu$_{24}$Sb$_{8}$Se$_{31}$) is shown in Fig.~\ref{fig.sevac.band}. In order to understand why a Se vacancy also behaves like a semiconductor we have to do a Se $p$ band count. In Cu$_{24}$Sb$_{8}$Se$_{32}$, there are 96 Se $p$ orbitals which split off into 8 lowest conduction bands and 88 $p$-valence bands. The 88 $p$-valence bands are filled up with 176 electrons (128 from Se, 24 from Cu and 24 from Sb). When we remove one Se atom, we take away 3 $p$ states. This removes 1 band from the 8 conduction bands and 2 bands from the $p$-valence bands, leaving 86 $p$-valence bands, which need 172 electrons to fill completely. Removal of one Se also removes 4 valence $p$ electrons giving exactly 172 valence electrons. One may ask what happen when we have more than one Se vacancies? Should the physics still be the same? Our calculations for two Se-vacancies interestingly reveal that they prefer different (SbSe$_4$) tetrahedrals, which means that our explanation above is still valid, at least for non-saturated concentration of Se-vacancies. In case of high concentration of Se-vacancies when there is chance of having two vacancies on the same (SbSe$_4$) tetrahedral, it will dope the system $n$-type rather than $p$-type. The explanation is similar to the case of one Se-vacancy, when removing the second Se, we remove three $p$ bands, but now all from valence bands, corresponding to six electrons, however, each Se have only four $p$ electrons that means there are two leftover electrons which will occupy the lowest of conduction bands and make the system $n$-type. Thus Se vacancy does not appear to be the source of $p$-type behavior of nominally pure Se4 as speculated by Skoug \etal\cite{Skoug.doping.2011}. We think the observed $p$-type behavior is most likely due to Cu vacancy as we have discussed earlier.

Replacing Se by S does not do much to the band structure because they belong to the same column whereas replacing Se ($s^2p^4$) by Cr ($s^2d^4$) completely alters the band structure (Fig.~\ref{fig.se.band}a,b). The semiconductor band structure is completely gone. One can take advantage of the electronic structure results for S by making solid solutions like Cu$_3$Sb$_{1-y}$Sn$_y$Se$_{1-x}$S$_x$ to reduce lattice thermal conductivity without affecting the optimized electronic transport properties. (see Skoug \etal\cite{skoug11.solidsolution}) 


To $n$-dope Cu$_3$SbSe$_4$ by substituting at the Se site we tried I ($s^2p^5$) and Mn ($s^2d^5$). Mn, like other transition metal atoms containing $d$ states changes the band structure dramatically and the system does not behave like a doped semiconductor (Fig.~\ref{fig.se.band}c). On the other hand, I appears to be good $n$-type dopant (Fig.~\ref{fig.se.band}c). However, as we discussed previously, due to the large positive formation energies, it is very difficult to dope Se4 on the Se site.
 

To test whether Cu$_3$SbSe$_4$ can be doped $p$-type by Se site substitution we tried Sb ($s^2p^3$) and V($s^2d^3$) substitution at the Se site (Fig.~\ref{fig.se.band}e,f). V, containing $d$ orbitals, destroys the semiconducting character and is not good. Sb$_{Se}$ also does not appear to be a good dopant.

\begin{figure}
\centering
\includegraphics[width=\columnwidth]{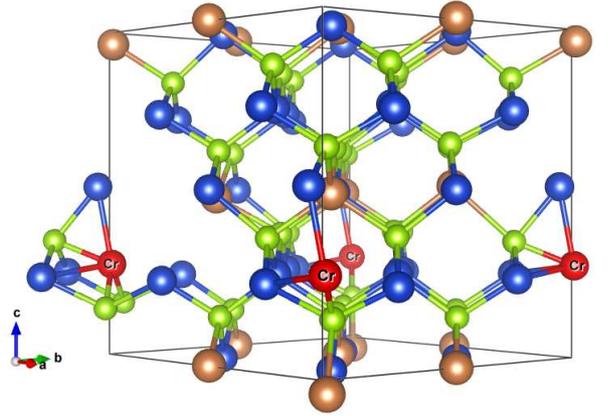}
\caption{\label{fig.cr_se_local_bond} Local bonds around Cr$_{Se}$ defect are distorted dramatically}
\end{figure}

It is noteworthy that while transition metals seem to do a good job in doping on the Sb site they perform poorly on the Se site. When substituting Se, transition metals change the electronic structure dramatically by introducing states in a wide range of energy, filling up the gap (Fig.~\ref{fig.se.band}). It turns out, the transition metals dramatically distort the local bondings, for example,  Fig.~\ref{fig.cr_se_local_bond} shows how Cr$_{Se}$ alters the local crystal structure. While a normal Se in Cu$_3$SbSe$_4$ coordinates to three Cu and one Sb, Cr$_{Se}$ moves away from the nn Sb and pulls closer to Se, totally opposite to the case of $Cr_{Sb}$ where the tetrahedral coordination is preserved. The difference between Sb and Se defect is the symmetry; Sb is at the center of a perfect tetrahedron of Se, on the other hand, Se is in a distorted tetrahedron of three Cu and on Sb. Such a change in local coordinations of transition metal defects on Se site also explains why their formation energies are large (Fig.~\ref{fig.formen}).

\section{\label{sec.sum}Summary and Conclusion}
Our work provides a systematic guide to search for good dopants in Cu$_3$SbSe$_4$. A combination of analyses of formation energy and band structure suggests that doping on Se site is not favorable (large positive formation energy), whereas it is easy to make $p$-type materials by substituting Sb with group $IV$ elements such as Sn, Ge, Pb and Ti, of which Ti defect may give better thermoelectric properties due to multiple hole pockets. There are just a few possible $n$-type dopants which include Mg and Zn substitution on the Cu site. The results also show that a possible cause for the observed $p$-type behavior of nominally pure Cu$_3$SbSe$_4$ is the presence of Cu vacancy rather than Se vacancy.
\section*{Acknowledgements}
This work was supported by the Center for Revolutionary Materials for Solid State Energy Conversion, an Energy Frontier Research Center funded by the U.S. Department of Energy, Office of Science, Office of Basic Energy Sciences under Award Number DE-SC0001054.

The calculations were done using computational resource provided by National Energy Research Scientific Computing Center(NERSC) and Michigan State University (MSU), Institution for Cyber Enabled Research (ICER), and High Performance Computer Center (HPCC). 

\bibliographystyle{elsarticle-num}
\bibliography{../../tex_reference/computational,../../tex_reference/thermoelectrics,../../tex_reference/tetrahedral,../../tex_reference/mypub}   

\begin{thebibliography}{10}
\expandafter\ifx\csname url\endcsname\relax
  \def\url#1{\texttt{#1}}\fi
\expandafter\ifx\csname urlprefix\endcsname\relax\def\urlprefix{URL }\fi
\expandafter\ifx\csname href\endcsname\relax
  \def\href#1#2{#2} \def\path#1{#1}\fi

\bibitem{pei_band_engin.2012}
Y.~Pei, H.~Wang, G.~J. Snyder, Band engineering of thermoelectric materials,
  Advanced Materials 24~(46) (2012) 6125--6135.
\newblock \href {http://dx.doi.org/10.1002/adma.201202919}
  {\path{doi:10.1002/adma.201202919}}.

\bibitem{Snyder2008.complex.thermo}
G.~Snyder, {Complex thermoelectric materials}, Nat. mater. 7~(2) (2008)
  105--114, and therein references.

\bibitem{sofo_mahan94}
J.~O. Sofo, G.~D. Mahan, Optimum band gap of a thermoelectric material, Phys.
  Rev. B 49 (1994) 4565--4570.
\newblock \href {http://dx.doi.org/10.1103/PhysRevB.49.4565}
  {\path{doi:10.1103/PhysRevB.49.4565}}.

\bibitem{ioffe57}
A.~F. Ioffe, Semiconductor Thermoelements and Thermoelectric Cooling, 1st
  Edition, Infosearch, 1957.

\bibitem{malsoon12}
M.-S. Lee, S.~D. Mahanti, Validity of the rigid band approximation in the study
  of the thermopower of narrow band gap semiconductors, Phys. Rev. B 85 (2012)
  165149.
\newblock \href {http://dx.doi.org/10.1103/PhysRevB.85.165149}
  {\path{doi:10.1103/PhysRevB.85.165149}}.

\bibitem{datahandbook}
O.~Madelung, Semiconductors: Data Handbook, 3rd Edition, Springer, 2004.

\bibitem{Skoug.doping.2011}
E.~Skoug, J.~Cain, P.~Majsztrik, M.~Kirkham, E.~Lara-Curzio, D.~Morelli,
  {Doping Effects on the Thermoelectric Properties of Cu$_3$SbSe$_4$}, Science
  of Advanced Materials 3~(4) (2011) 602--606.

\bibitem{skoug11.solidsolution}
E.~J. Skoug, J.~D. Cain, D.~T. Morelli, High thermoelectric figure of merit in
  the {Cu$_3$SbSe$_4$}-{Cu$_3$SbS$_4$} solid solution, Appl. Phys. Lett.
  98~(26) (2011) 261911.

\bibitem{biswas_high-performance_2012}
K.~Biswas, J.~He, I.~D. Blum, C.-I. Wu, T.~P. Hogan, D.~N. Seidman, V.~P.
  Dravid, M.~G. Kanatzidis, High-performance bulk thermoelectrics with
  all-scale hierarchical architectures, Nature 489~(7416) (2012) 414--418.
\newblock \href {http://dx.doi.org/10.1038/nature11439}
  {\path{doi:10.1038/nature11439}}.

\bibitem{skoug_thesis_2011}
E.~J. Skoug, Copper-based diamond-like ternary semiconductors for
  thermoelectric applications, Ph.d., Michigan State University, United States
  -- Michigan (2011).

\bibitem{do_cusbse.2012}
D.~Do, V.~Ozolins, S.~D. Mahanti, M.-S. Lee, Y.~Zhang, C.~Wolverton, Physics of
  bandgap formation in {Cu}-–{Sb}-–{Se} based novel thermoelectrics: the
  role of {Sb} valency and cu $d$ levels, J. Phys.: Condens. Matter 24~(41)
  (2012) 415502.

\bibitem{do_bond_band2014}
D.~T. Do, S.~Mahanti, Bonds, bands, and band gaps in tetrahedrally bonded
  ternary compounds: The role of group {V} lone pairs, Journal of Physics and
  Chemistry of Solids 75~(4) (2014) 477 -- 485.
\newblock \href
  {http://dx.doi.org/http://dx.doi.org/10.1016/j.jpcs.2013.12.004}
  {\path{doi:http://dx.doi.org/10.1016/j.jpcs.2013.12.004}}.

\bibitem{li_Bi_Sb.2013}
X.~Li, D.~Li, H.~Xin, J.~Zhang, C.~Song, X.~Qin, Effects of bismuth doping on
  the thermoelectric properties of {Cu$_3$SbSe$_4$} at moderate temperatures,
  Journal of Alloys and Compounds 561~(0) (2013) 105 -- 108.
\newblock \href
  {http://dx.doi.org/http://dx.doi.org/10.1016/j.jallcom.2013.01.131}
  {\path{doi:http://dx.doi.org/10.1016/j.jallcom.2013.01.131}}.

\bibitem{wei_Cu_deficiency.2014}
T.-R. Wei, F.~Li, J.-F. Li, Enhanced thermoelectric performance of
  nonstoichiometric compounds {Cu$_{3-−x}$ SbSe$_4$} by {Cu} deficiencies,
  Journal of Electronic Materials 43~(6) (2014) 2229--2238.
\newblock \href {http://dx.doi.org/10.1007/s11664-014-3018-4}
  {\path{doi:10.1007/s11664-014-3018-4}}.

\bibitem{wei_Sn_Sb.2014}
T.-R. Wei, H.~Wang, Z.~M. Gibbs, C.-F. Wu, G.~J. Snyder, J.-F. Li,
  Thermoelectric properties of {Sn}-doped p-type {Cu$_3$SbSe$_4$}: a compound
  with large effective mass and small band gap, J. Mater. Chem. A 2 (2014)
  13527--13533.
\newblock \href {http://dx.doi.org/10.1039/C4TA01957A}
  {\path{doi:10.1039/C4TA01957A}}.

\bibitem{zhang_northrup.91}
S.~B. Zhang, J.~E. Northrup, Chemical potential dependence of defect formation
  energies in gaas: Application to ga self-diffusion, Phys. Rev. Lett. 67
  (1991) 2339--2342.
\newblock \href {http://dx.doi.org/10.1103/PhysRevLett.67.2339}
  {\path{doi:10.1103/PhysRevLett.67.2339}}.

\bibitem{bloch94}
P.~E. Bl\"ochl, Projector augmented-wave method, Phys. Rev. B 50 (1994)
  17953--17979.

\bibitem{kresse99}
G.~Kresse, D.~Joubert, From ultrasoft pseudopotentials to the projector
  augmented-wave method, Phys. Rev. B 59~(3) (1999) 1758--1775.

\bibitem{pbe}
J.~P. Perdew, K.~Burke, M.~Ernzerhof, Generalized gradient approximation made
  simple, Phys. Rev. Lett. 77~(18) (1996) 3865--3868.

\bibitem{vasp1}
G.~Kresse, J.~Hafner, Ab initio molecular dynamics for liquid metals, Phys.
  Rev. B 47~(1) (1993) 558--561.

\bibitem{vasp2}
G.~Kresse, J.~Furthm\"uller, Efficiency of ab-initio total energy calculations
  for metals and semiconductors using a plane-wave basis set, Comput. Mater.
  Sci. 6~(1) (1996) 15 -- 50.

\bibitem{vasp3}
G.~Kresse, J.~Furthm\"uller, Efficient iterative schemes for ab initio
  total-energy calculations using a plane-wave basis set, Phys. Rev. B 54~(16)
  (1996) 11169--11186.

\bibitem{martin_book}
R.~M. Martin, Electronic Structure: Basic Theory and Practical Methods,
  Cambridge University Press, 2004.

\bibitem{hse06:1}
J.~Heyd, G.~E. Scuseria, M.~Ernzerhof, Hybrid functionals based on a screened
  {C}oulomb potential, J. Chem. Phys. 118 (2003) 8207.

\bibitem{hse06:2}
J.~Heyd, G.~E. Scuseria, Efficient hybrid density functional calculations in
  solids: Assessment of the {Heyd}-–{Scuseria}-–{Ernzerhof} screened
  {Coulomb} hybrid functional, J. Chem. Phys. 121 (2004) 1187.

\bibitem{hse06:3}
J.~Heyd, G.~E. Scuseria, M.~Ernzerhof, Erratum: {“Hybrid} functionals based
  on a screened {Coulomb} potential” {[J.} chem. phys. 118, 8207 (2003)], J.
  Chem. Phys. 124 (2006) 219906.

\bibitem{dudarev98}
S.~L. Dudarev, G.~A. Botton, S.~Y. Savrasov, C.~J. Humphreys, A.~P. Sutton,
  Electron-energy-loss spectra and the structural stability of nickel oxide:
  {An} {LSDA+U} study, Phys. Rev. B 57 (1998) 1505.

\bibitem{monkhorst76}
H.~J. Monkhorst, J.~D. Pack, Special points for brillouin-zone integrations,
  Phys. Rev. B 13~(12) (1976) 5188--5192.

\bibitem{defects}
R.~M. Nieminen, Topics in Applied Physics: Theory of defects in semiconductors,
  Vol. 104, Springer, 2006, pp. 36--40.

\bibitem{chistensen_correction.1984}
N.~E. Christensen, Electronic structure of gaas under strain, Phys. Rev. B 30
  (1984) 5753--5765.
\newblock \href {http://dx.doi.org/10.1103/PhysRevB.30.5753}
  {\path{doi:10.1103/PhysRevB.30.5753}}.

\bibitem{segev_self_corrections.2007}
D.~Segev, A.~Janotti, C.~G. Van~de Walle, Self-consistent band-gap corrections
  in density functional theory using modified pseudopotentials, Phys. Rev. B 75
  (2007) 035201.
\newblock \href {http://dx.doi.org/10.1103/PhysRevB.75.035201}
  {\path{doi:10.1103/PhysRevB.75.035201}}.

\bibitem{suzumura_s4_2014}
A.~Suzumura, M.~Watanabe, N.~Nagasako, R.~Asahi, Improvement in thermoelectric
  properties of se-free cu3sbs4 compound, Journal of Electronic Materials
  43~(6) (2014) 2356--2361.
\newblock \href {http://dx.doi.org/10.1007/s11664-014-3064-y}
  {\path{doi:10.1007/s11664-014-3064-y}}.

\bibitem{Heremans_distorted_dos.2008}
J.~P. Heremans, V.~Jovovic, E.~S. Toberer, A.~Saramat, K.~Kurosaki,
  A.~Charoenphakdee, S.~Yamanaka, G.~J. Snyder, Enhancement of thermoelectric
  efficiency in pbte by distortion of the electronic density of states, Science
  321~(5888) (2008) 554--557.
\newblock \href {http://dx.doi.org/10.1126/science.1159725}
  {\path{doi:10.1126/science.1159725}}.

\end{thebibliography}
\end{document}